\pdfoutput=1
\documentclass[twocolumn,nofootinbib,preprintnumbers,amsmath,amssymb]{revtex4}
\usepackage{amssymb}
\usepackage{graphicx}
\usepackage{epsfig}

\begin{document}

%\preprint{MCTP-11-} 

\title{Higgs Mass Prediction for Realistic String/$M$ Theory Vacua}
\author{Gordon Kane$^{\dag}$} 
\author{Piyush Kumar$^{\star}$}
\author{Ran Lu$^{\dag}$}
\author{Bob Zheng$^\dag$}

\affiliation{$^\dag$Michigan Center for Theoretical Physics, University of Michigan, Ann Arbor, MI 48109 USA \\ \\
$^\star$Department of Physics $\&$ ISCAP, Columbia University, New York, NY 10027 USA}

\date{\today}
% It is always \today, today
%  but any date may be explicitly specified

\begin{abstract}
Recently it has been recognized that in compactified
string/$M$ theories that satisfy cosmological constraints, it is possible to
derive some robust and generic predictions for particle physics and cosmology with very mild assumptions. 
When the matter and gauge content below the compactification scale is that of the MSSM, it is possible to make precise predictions. 
In this case, we predict that there will be a single Standard Model-like Higgs boson with a
calculable mass 105 GeV $\lesssim M_{h}\lesssim 129$ GeV depending on $\tan\beta $ (the ratio of the Higgs
vevs in the MSSM). For $\tan{\beta} >7$, the prediction is : 122 GeV $\lesssim  M_h \lesssim$ 129 GeV.
\end{abstract}

%\pacs{11.25.Mj 11.25.Wx 11.25.Yb 12.10.-g 12.60.Jv 14.80.Ly}% PACS, the Physics and Astronomy
% Classification Scheme.
%\keywords{Suggested keywords}%Use showkeys class option if keyword
%display desired

\maketitle

%\tableofcontents

%\newpage

\section{Motivation}\label{motivation}

Most physicists agree that understanding the origin of electroweak
symmetry breaking is essential for progress in going
beyond the Standard Model. The LHC experiments have made tremendous progress in constraining the
Higgs mass in the past year or so.
The combined results from the LEP, the TeVatron and the LHC will soon cover the entire
region below about $500$ GeV. We will demonstrate that, 
with some broad and mild assumptions motivated by cosmological constraints, 
generic compactified string/$M$-theories with stabilized moduli and low-scale supersymmetry imply a
Standard Model-like single Higgs boson with a mass $105\,{\rm GeV} \lesssim
M_{h}\lesssim 129\,{\rm GeV}$ if the matter and gauge spectrum surviving below the compactification scale is 
that of the MSSM, as seen from Figure 1. 
%The upper limit may be a little larger or smaller depending on the 
%gravitino mass which sets the scale for supersymmetry breaking, as also seen from the Figure. 
For an extended gauge and/or matter spectrum, there can be additional contributions to $M_{h}$.  
Furthermore, in  
$G_2$-MSSM models   \cite{arXiv:0801.0478} we find that the range of possible Higgs masses is apparently much smaller,  
$122\,{\rm GeV} \lesssim M_{h}\lesssim 129\,{\rm GeV}$.

The Standard Model suffers from ``naturalness" or "hierarchy" problem(s).  In addition to the well-known technical naturalness problem
of the Higgs, there is the basic question of the origin of the electroweak scale. In the context considered here:
the embedding of the (supersymmetric)
Standard Model in a UV complete microscopic theory like string/$M$ theory has
to explain why the electroweak scale is so much smaller than the natural
scale in string theory, the string scale, which is usually assumed to be many of orders of magnitude above the 
TeV scale. The $\mu $ parameter (which sets the masses of Higgsinos and
contributes to the masses of Higgs bosons) must also be around TeV scale. 
The models we describe here, with
softly broken supersymmetry, include solutions for all of these problems.

\begin{figure}[h!]
\includegraphics[width=3.45in,height=2.9in]{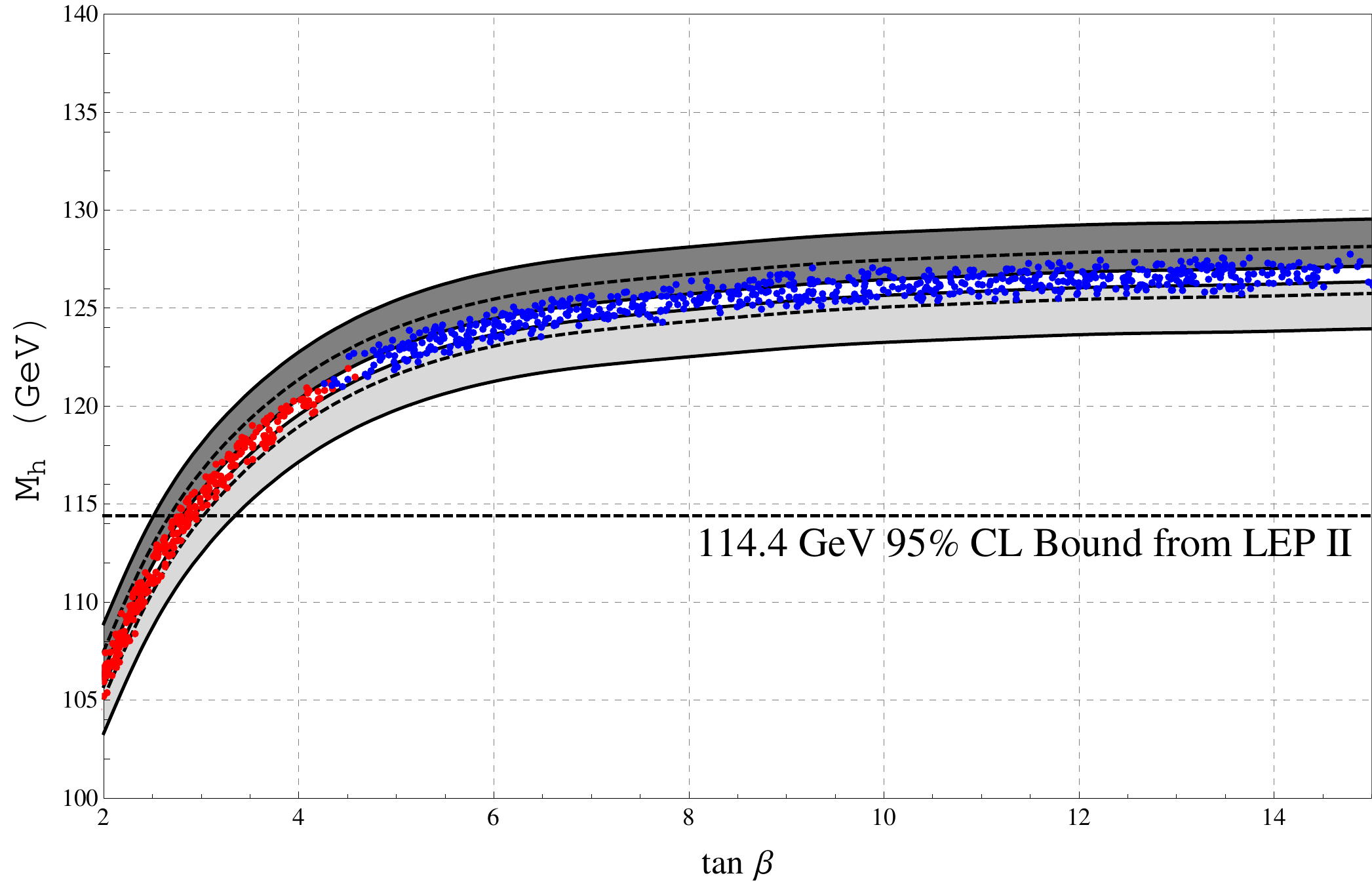}
\caption{\small{The prediction for the Higgs mass at two-loops for realistic string/$M$ theory vacua defined in the text, as a function of $\tan \beta$ for three different values of the gravitino mass $m_{3/2}$, and varying the theoretical and experimental inputs as described below. For precise numbers and more details, see section \ref{results}. The central band within the dashed curves for which scatter points are plotted corresponds to $m_{3/2}=50$ TeV. This band includes the total uncertainty in the Higgs mass arising from the variation of three theoretical inputs at the unification scale, and from those in the top mass $m_t$ and the $SU(3)$ gauge coupling $\alpha_s$ within the allowed uncertainties. The innermost (white) band bounded by solid curves includes the uncertainty in the Higgs mass for $m_{3/2}=50$ TeV only from theoretical inputs. The upper (dark gray) band bounded by solid curves corresponds to the total uncertainty in the Higgs mass for $m_{3/2}=100$ TeV while the lower (light gray) band bounded by solid curves  corresponds to that for $m_{3/2}=25$ TeV.  For $m_{3/2}=50$ TeV, the red scatter points (with $\tan\beta$ less than about 4.5) and blue scatter points (with $\tan\beta$ greater than about 4.5) correspond to ``Large" $\mu$ and ``Small" $\mu$ respectively, as described in section \ref{Higgs-general} and section \ref{results}.}}
\end{figure}

Although understanding phenomenologically relevant supersymmetry breaking in string theory is a challenging task, 
many results, including those needed to calculate the Higgs
boson mass, can be obtained with rather mild, well motivated assumptions. The rest of this section
outlines and motivates these simple assumptions.

In connecting string/$M$ theory to low-energy particle physics, one has to
compactify the extra dimensions. Motivated by grand unification and its successful embedding into string/$M$ theory
we assume that the
string/$M$ scale, the Kaluza-Klein scale and the unification scale are all within an order of magnitude of
$10^{16}$ GeV. 
Within the theoretical precision desired, numerical results for $M_{h}$ are not
sensitive to variations of an order of magnitude or so in these scales. 

Even though the energy scale of the extra dimensions is assumed to be much above the center-of-mass energy of
collisions at the LHC, the extra dimensions still manifest themselves at lower energies through the presence of ``moduli"
fields. These are modes of the extra dimensional graviton whose vacuum-expectation-values (\emph{vev}'s) determine the shapes and sizes that
the extra dimensions take. Being modes of the extra dimensional graviton, the moduli couple to matter with Planck suppressed
interactions universally. The moduli have to be stabilized since all couplings and masses are determined
from their \emph{vev}'s.

In recent years, significant progress has been made in understanding 
moduli stabilization and supersymmetry breaking in different
corners of string/$M$ theory, see  
\cite{hep-th/0212294, hep-th/0301240, hep-th/0503124, arXiv:hep-th/0502058,arXiv:1003.1982,hep-th/0606262,hep-th/0701034,hep-th/0505160,arXiv:1002.1081,arXiv:1102.0011}. 
In this work, we will be interested in
supersymmetry breaking mechanisms which give rise
to TeV-scale supersymmetry, and hence solve the
naturalness problems in the Standard Model. The basic mechanisms were
described in \cite{hep-th/0606262,hep-th/0701034} for $M$-theory and in \cite{hep-th/0301240,hep-th/0503124,arXiv:1003.1982} for Type IIB compactifications,
where it was shown that all moduli can be stabilized and supersymmetry can
be broken with $\sim $ TeV-scale superpartners with a natural choice of
parameters - in which the only dimensionful scale is $M_{pl}$! In a vacuum with broken supersymmetry and vanishing cosmological constant, the mass of the gravitino ($m_{3/2}$), which is the 
superpartner of the massless graviton, is the order-parameter of supersymmetry breaking and sets the mass scale for all superpartners and also indirectly the Higgs mass.

A natural prediction of such compactifications is that the mass of the lightest modulus
is close to $m_{3/2}$.
In fact, this is a generic result\ for compactified string/$M$-theories with
stabilized moduli\ within the supergravity approximation. In vacua
in which the superpotential $W$ is not tuned, it essentially arises from the
fact that there is a relationship between the dynamics
stabilizing the moduli and the dynamics breaking supersymmetry due to the
extremely tiny value of the cosmological constant. 
Thus, the modulus mass
becomes related to the gravitino mass.  For more details, see \cite{hep-th/0611183}. For the generic case with 
many moduli, at least some of the moduli are stabilized by perturbative effects in the K\"{a}hler potential \cite{hep-th/0701034,arXiv:1003.1982}. Then, it can be shown that the lightest modulus\footnote{
more precisely, the real part of the chiral superfield making up the complex
modulus in the 4D theory.} has a mass of the same order as $m_{3/2}$ \cite{hep-th/0602246,arXiv:1006.3272}.
%Technically this occurs because the scalar goldstino partners of the
%goldstinos that become the spin $1/2$ polarizations of gravitinos when
%supersymetry is broken, mix with moduli and thus enter the moduli mass
%matrix.  Even though the moduli mass matrix is not known in much detail,
%one can show it has one or more eigenvalues less than or about equal to the
%sgoldstino mass, which is generically about $m_{3/2}$.

\subsection{Generic Predictions}\label{generic}  

The fact that, generically, the lightest modulus mass is of the same order as the gravitino mass has significant implications
for the phenomena described by a typical string/$M$ theory vacuum, with some rather mild assumptions.  In addition to the requirement of stabilizing all moduli in a vacuum with TeV-scale supersymmetry (as described in the previous section) which picks the set of string compactifications we choose to study, the assumptions are essentially that the supergravity
approximation is valid and that the Hubble scale during inflation is larger
than $m_{3/2}$. 

In particular, the above implies that the light moduli fields (of order $m_{3/2}$) are generically displaced during inflation, causing the Universe to become moduli-dominated shortly after the end of inflation due to coherent oscillations of the moduli. Requiring that these decay before big-bang nucleosynthesis (BBN) so as to not ruin its predictions, puts a lower bound on $m_{3/2}$ of about 25 TeV or so.  Thus, the cosmological moduli problem \cite{CMP} generically requires that $m_{3/2} > 25$ TeV.
Since $m_{3/2}\simeq\frac{F}{M_{pl}}$ for vacua with a vanishingly small cosmological constant, 
this further implies that the supersymmetry breaking scale $\sqrt{F
}$ has to be ``high", as is natural in gravity mediation
models. Low-scale supersymmetry breaking scenarios like gauge mediation do
not seem to be compatible with these cosmological constraints. 

As an aside,
the requirement of stabilizing a large number of moduli in a realistic
compactification with a simple mechanism naturally picks mechanisms in which
many axions\footnote{the imaginary parts of the complex moduli} are exponentially lighter than $%
m_{3/2}$, one of which can naturally be the QCD axion \cite{arXiv:1004.5138,arXiv:1003.1982}. Hence, this
provides a string theory solution of the strong-CP problem with stabilized
moduli and axions, and also naturally predicts an $\mathcal{O}(1)$ fraction
of Dark Matter in the form of axions, the abundance of which must now be computed
with a non-thermal cosmological history \cite{arXiv:1004.5138}.

Generically, within high scale supersymmetry breaking mechanisms such as gravity mediation, 
the squark, slepton and heavy Higgs masses are also of order $m_{3/2}$. It has been argued that in special cases, 
squarks and sleptons may be ``sequestered" from supersymmetry breaking, giving rise to a suppression in their masses relative to $m_{3/2}$. 
However, it was shown in \cite{arXiv:1012.1858} that in
string/$M$ theory compactifications with moduli stabilization, the
squark and slepton masses are generically not sequestered from supersymmetry
breaking once all relevant effects are taken into account.
This has important implications for collider physics, implying in particular
that squarks and sleptons should \emph{not} be directly observed at the LHC \cite{arXiv:1006.3272,arXiv:0901.3367}.

While scalar superpartner masses are tied to $m_{3/2}$, gaugino masses need not be.
Within many classes of string compactifications which
satisfy all the requirements stated above, it can be shown that the
gaugino masses \cite{hep-ph/0511162, arXiv:0801.0478} and $\mu$ \cite{arXiv:1102.0556} are suppressed by one-to-two orders of magnitude relative to $m_{3/2}$. 
In this case, it can be shown that the LSP could naturally provide the DM abundance with a
non-thermal mechanism \cite{arXiv:0804.0863,arXiv:0908.2430}. However, it is not clear at present if this is a generic
feature of \emph{all} realistic compactifications. For example, it is possible in classes of 
string compactifications to stabilize moduli in such 
a way that the gaugino masses are of the same order as $m_{3/2}$ \cite{arXiv:0805.1029}. 
Similarly, it is possible for $\mu$ to be generated at the same order as $m_{3/2}$ \cite{hep-ph/9302227} .
A review of particle physics and cosmology in this general  framework will apear shortly in \cite{AKK}.

\section{The Higgs and BSM Physics}\label{Higgs-general}

We are interested in making predictions for the Higgs mass
arising from realistic compactifications satisfying the conditions above. 
In a supersymmetric theory, two Higgs doublets
are required for anomaly cancellation; so by the ``Higgs
mass" we mean the mass of the lightest CP-even neutral scalar in the Higgs
sector. A remarkable fact about the Higgs mass in
general supersymmetric theories is that an upper limit on $M_h$ of order $2\,M_{Z}$
exists just from the requirement of validity of perturbation
theory up to the high scale of order $10^{16}$ GeV \cite{hep-ph/9210242}. 
This is due to the
fact that the Higgs mass at tree-level only depends on SM gauge couplings
(which have been measured), and possibly other Yukawa or gauge couplings
(which are bounded from above by perturbativity). However, in addition to
the gauge and matter spectrum, the precise value of the Higgs mass depends
crucially on radiative effects, which in turn depend on all the soft
parameters including the $\mu $ and $B\mu $ parameters.

In this work we assume that the visible sector consists of the SM gauge
group with the MSSM matter content below the unification scale, as suggested
by gauge coupling unification and radiative EWSB in the MSSM. In addition, we consider
compactifications in which the gravitino mass $m_{3/2}$ is not too far above
the lower bound of $\sim 25$ TeV from the moduli decay constraint, and the
gaugino masses  are suppressed by one-to-two orders
of magnitude relative to $m_{3/2}$. For $\mu$ we study two cases, one in which $\mu$ is suppressed 
by one-to-two orders of magnitude relative to $m_{3/2}$ as predicted in \cite{arXiv:1102.0556}, and the other in which $\mu$ is of the same order
as $m_{3/2}$  \cite{hep-ph/9302227}. We denote these two cases as ``Small" $\mu$ and ``Large" $\mu$ respectively. The two cases are studied as they pick out different regions of $\tan \beta$ and hence give different predictions for the Higgs mass as seen from Figure 1. 
For more discussion, see section \ref{results}.

Note that only one of the five scalars in the Higgs sector of the MSSM is light, the rest are all of
order $m_{3/2}$. Hence, we are in the ``decoupling limit" of
the MSSM where the lightest CP-even Higgs scalar has precisely the same
properties as the SM Higgs\footnote{%
In the decoupling limit, the Higgs mixing angle denoted by $\alpha $ in \cite{hep-ph/9709356}
is given by $\alpha =\beta -\frac{\pi }{2}$, where $\beta \equiv \tan ^{-1}(%
\frac{v_{u}}{v_{d}})$. $v_{u}$ and $v_{d}$ are the vacuum-expectation-values
(\emph{vevs}) of the two neutral Higgs fields $H_{u}^{0}$ and $H_{d}^{0}$ in
the MSSM.}. The low-energy theory arising from $M$-theory studied in \cite{arXiv:0801.0478} 
naturally gives rise to these features, but the results apply to all compactifications with
scalars heavier than about 25 TeV and $\lesssim$ TeV gauginos. The Higgs mass can then be reliably computed with a small and
controlled theoretical uncertainty. This will be the subject of the
following sections.

From a bottom up point of view some authors have noted that heavy scalars have some attractive features \cite{hep-ph/9607394}  
and related phenomenology has been studied in \cite{arXiv:1109.3197, arXiv:1105.3765}. Their
conclusions are consistent with ours where they overlap. The framework considered here is quite different from split supersymmetry \cite{hep-th/0405159} 
and high-scale supersymmetry \cite{arXiv:0910.2235} which have much heavier scalars.  In split supersymmetry
\emph{both} gaugino masses and trilinears are suppressed relative to scalars by a symmetry -- in this case an $R$-symmetry, 
together with supersymmetry breaking of the $D$-type as described in \cite{hep-th/0405159}. In contrast, in the class of compactifications considered here,  
the gaugino masses are suppressed by dynamics, since the $F$-term for the modulus determining the gaugino masses is suppressed relative to the dominant $F$-term. 
Hence, gaugino masses can only be suppressed by one-to-two orders of magnitude, not arbitrarily as in split-supersymmetry. 
For the same reason, the gluinos are not ``long-lived" in the realistic string/$M$ theory vacua under consideration here. 
Also, the trilinears are not suppressed at all. 
With large trilinears, one has to be careful about charge and 
color breaking (CCB) minima, and we have confirmed the absence of these in models of interest. Another notable difference from
split-supersymmetry and high-scale supersymmetry is that, in those models,
(radiative) electroweak symmetry breaking is not implemented when computing the Higgs mass 
since a huge fine-tuning is present \emph{by assumption}. In contrast, in the
string/$M$ theory models considered in this work,  (radiative) electroweak
symmetry breaking occurs naturally in a large subset of the parameter space. 
However, the ease in obtaining the correct value of the Higgs vev (or $Z$-boson mass) 
depends on the value of $\mu$. For ``Small" $\mu$, it can be shown that
the fine-tuning involved in obtaining the correct Higgs vev is significantly
reduced compared to the naive expectation for heavy scalars due to an automatic cancellation between scalar masses and
trilinears which are both close to $m_{3/2}$ in this setup; for details see
\cite{arXiv:1105.3765}. This can naturally give rise to $\mu \lesssim $ TeV, even when the
scalar mass parameters are $\gtrsim 30$ TeV. For ``Large" $\mu$, the fine-tuning is quite severe as one would expect.
We include both cases here.

\section{Computation of the Higgs Mass}\label{compute}

Computing the Higgs mass in the MSSM with scalar masses and trilinears at $%
M_{susy}\gtrsim $ 25 TeV, and gauginos and $\mu $ suppressed by one-to-two orders of
magnitude relative to the scalar masses, is non-trivial. Although conceptually quite different, some of the
technical issues involved have an overlap with split-supersymmetry 
and high-scale supersymmetry.

Since the scalar masses are much larger than a TeV, they could lead to
non-trivial quantum corrections in the gaugino-higgsino and Higgs sectors
enhanced by \textquotedblleft large logarithms" of the ratio between the
electroweak scale and the scalar mass scale. Many numerical codes tend
to become less reliable for scalar masses larger than a few TeV for the
above reason. However, in contrast to split supersymmetry and high-scale
supersymmetry models, the scalar masses here are $25$-$100$ TeV which is not 
that ``large", since $\log (\frac{M_{susy}}{M_{EW}})$ is not
large. So, numerical codes should still provide a reasonable estimate. \
The ratio of the two Higgs fields vevs, $\tan\beta $ cannot yet be calculated
accurately, and significantly affects the value of $M_{h}$ if $\tan\beta
\lesssim 10$, so we include the variation from $\tan\beta$. This dependence
actually allows an approximate measurement or useful limit on $\tan\beta$
which is otherwise very difficult to do.

In light of the above, we adopt the following procedure. At the unification
scale around $10^{16}$ GeV, in accord with theoretical expectations we fix the soft parameters - the scalar
masses equal to $m_{3/2}$, the trilinears $A$ close to $m_{3/2}$,
and the gaugino masses suppressed by one-to-two orders of magnitude
relative to the scalar masses as described in \cite{arXiv:0801.0478}. Then, for a given value
of $\tan \beta $, the numerical codes SOFTSUSY \cite{Allanach:2001kg} and SPHENO \cite{Porod:2003um} are used to renormalize these quantities 
down to $M_{susy}\approx m_{3/2}$, where electroweak symmetry breaking is
implemented. This determines $\mu $ and $B\mu $. The quantities are chosen such that the
values of $\mu $ and $B\mu $ are consistent with the theoretical
expectations. One consequence of this is that $\tan\beta $ is not
expected to span the fully phenomenologically allowed range from $\sim 2$ to $\sim 60$,
but only a restricted range from $\sim 2$ to $\sim 15$ \cite{arXiv:1006.3272}. In any case, from Figure 1, since the Higgs mass saturates for 
$\tan\beta \gtrsim 12$, plotting higher values of $\tan\beta$ will not provide new information.

Then, we compute the Higgs mass in the ``match-and-run" approach using
values of gaugino masses, $\mu $ and $B\mu $ at $M_{susy}$ determined from
above. We follow a procedure similar to that in
\cite{arXiv:1108.6077} except that we only consider those parameters at the unification 
scale which after RG evolution to $M_{susy}$ give rise to viable electroweak symmetry breaking. We also compute the Higgs mass directly 
with SOFTSUSY using theoretical inputs at the unification scale, and compare to the results obtained with the ``match-and-run" approach, the detailed procedure for which is described  below.

\subsection{Matching at $M_{susy}$}\label{matching-msusy}

At the scale $M_{susy}$, the full supersymmetric theory is matched to a low
energy theory with fewer particles, consisting of the SM particles, the gauginos and the higgsinos for the ``Small" $\mu$ case, and 
only the SM particles and the gauginos for the ``Large" $\mu$ case. The
matching condition for the quartic coupling of the Higgs in the low-energy
theory is given at $M_{susy}$ by: 
\begin{eqnarray}
\lambda =\frac{1}{4}\left[g_2^2+\frac{3}{5}\,g_1^2\right]\,\cos^2\,2\beta +
\delta_{th}^{\lambda}
\end{eqnarray}
where $g_1,\,g_2$ are the $U(1)_Y$ and $SU(2)_L$ gauge couplings evaluated
at $M_{susy}$. The threshold corrections to the quartic coupling at one-loop consist
of leading log (LL) as well as finite corrections. The above matching condition is strictly valid only in the $\overline{DR}$ scheme, so there is an 
additional correction if one wants to convert to the $\overline{MS}$ scheme as explained in the appendix of \cite{arXiv:0705.1496}. 
We use the standard choice $M_{susy} = \sqrt{M_{\tilde{t}_1}\,M_{\tilde{t}_2}}$ where $M_{\tilde{t}_1},M_{\tilde{t}_2}$ are the 
masses of the two stop squarks, and include all the relevant LL and finite threshold corrections. 
The dominant finite threshold effects to the Higgs quartic coupling comes from stop squarks, and is given by: 
\begin{eqnarray}
\delta_{th}^{\lambda} \approx \frac{3\,y_t^4}{8\pi^2}\,\left(\frac{A_t^2}{m_{%
\tilde{t}}^2}-\frac{A_t^4}{12\,m_{\tilde{t}}^4}\right)
\end{eqnarray}
Since the trilinears $A$ are of the same order as scalars, this is
a non-trivial correction when the scalars and trilinears are around 25 TeV.
Other finite threshold corrections are smaller, and have been neglected as they do not affect the result to within the accuracy desired.

The matching conditions for the gaugino-higgsino-Higgs couplings (denoted by $\kappa$ in general) at $M_{susy}$ in the $\overline{DR}$ scheme are given by: 
\begin{eqnarray}\label{kappa}
\kappa_{2u}&=&g_2\,\sin\,\beta;\;\kappa_{2d}=g_2\,\cos\,\beta;\; 
\nonumber \\
\kappa_{1u}&=&\sqrt{\frac{3}{5}}\,g_1\,\sin\,\beta;\;\kappa_{1d}=\sqrt{%
\frac{3}{5}}\,g_1\,\cos\,\beta;\;
\end{eqnarray}
where the gauge couplings are to be evaluated at $M_{susy}$. As for the Higgs quartic coupling, additonal corrections are present in the $\overline{MS}$
scheme, which can be obtained from \cite{arXiv:0705.1496}.

\subsection{Two -loop RGEs and Weak Scale Matching}\label{matching-weak}

We use two-loop RGEs computed in \cite{arXiv:1108.6077} for the gauge couplings,
third-generation Yukawa couplings $y_t,\,y_b,\,y_{\tau}$, the Higgs
quartic $\lambda$, and the
gaugino-higgsino-Higgs ($\kappa$) couplings (for ``Small" $\mu$),  to renormalize them down to the weak scale. 
For ``Large" $\mu$, the $\kappa$ couplings are not present in the low-energy theory, and (\ref{kappa}) is used to compute the threshold 
correction from higgsinos at $M_{susy}$.

Note that unlike the $\kappa$ couplings and the quartic coupling, the
boundary conditions for which are defined at $M_{susy}$, the boundary
conditions for the gauge and Yukawa couplings $y_b, y_{\tau}$ are defined at 
$M_Z$ - the $Z$-pole, while that for the top Yukawa coupling $y_t$ is
defined at the top pole mass $m_{t}= 173.1\pm 0.9$ GeV \cite{arXiv:1107.5255}. In particular,
the boundary values of the running gauge and Yukawa couplings in the $
\overline{MS}$ scheme are extracted from experimental observables at the
weak scale by including threshold effects, as explained in \cite{arXiv:0705.1496}. 
For the top Yukawa coupling $y_t$, non-trivial three-loop QCD
corrections, and one-loop electroweak and superpartner threshold corrections
are also included as they are non-negligible and play an important role in the precise prediction
for the Higgs mass.Since the boundary conditions are given at different scales, an iterative
procedure is required to solve the coupled differential RGE equations. We
follow a procedure similar to that in \cite{arXiv:1108.6077, hep-ph/0408240,arXiv:0705.1496}. Then, the Higgs mass is given by:
\begin{equation}
M_h = \sqrt{2}\,v\,\sqrt{\lambda(Q) + \delta_{\lambda} (Q) + \tilde{\delta}_{\lambda} (Q)}
\end{equation}
where $v=174.1$ GeV, $\delta_{\lambda}$ stands for the corrections from the SM particles, and $\tilde{\delta}_{\lambda}$ stands for the corrections from the supersymmetric fermions at the weak scale, and all couplings are evaluated at the $\overline{MS}$ scale $Q = m_t$. The expressions are given in \cite{arXiv:1108.6077}. Finally, as mentioned earlier, since numerical codes are expected to give a good estimate of the Higgs mass, we compute the Higgs mass directly with SOFTSUSY. We find very good agreement between the two results, to within 1 GeV. 

\section{Result}\label{results}

Figure 1 gives the Higgs mass as a function of $\tan\beta$ by varying the theoretical inputs at the unification scale in ranges predicted 
by the theory, and $m_t$ and $\alpha_s$ within the allowed uncertainties. The values of the Higgs mass shown are in the ``match-and-run" approach. The ranges of the theoretical and experimental inputs,  and the resulting uncertainties are discussed in detail below. 
The $\mu$ and $B\mu$ parameters are related by electroweak symmetry breaking to $\tan\beta$, so by varying $\tan\beta$ one is effectively varying $\mu$ and $B\mu$. 
As pointed out in section \ref{motivation}, 
theoretical considerations typically give rise to two different classes of phenomenologically viable predictions for $\mu$ -- 
one in which $\mu$ is suppressed by one-to-two orders of magnitude relative to $m_{3/2}$, and the other in which $\mu$ is comparable to $m_{3/2}$. As seen from Figure 1, the two classes of predictions for $\mu$ give rise to different values of $\tan\beta$ because of the EWSB constraints that relates them; 
hence a measurement of the Higgs mass will not only determine or constrain $\tan \beta$, it will also favor one class of $\mu$-generating mechanisms over the other! 
For instance, in $G_2$-MSSM models arising from $M$ theory, Witten's solution to the doublet-triplet splitting problem \cite{hep-ph/0201018}
results in $\mu$ being suppressed by about an order of magnitude. Hence, in these vacua, the Higgs mass sits in the range 122 GeV $\lesssim M_h \lesssim$ 129 GeV.

{\renewcommand{\arraystretch}{1.5}
\begin{table}[h!]
\begin{tabular}{| p{3.4cm} | p{3.9 cm} |p{1.2cm}|}
\hline
Case & Variation of Input & $\Delta M_h$ \\ 
\hline
``Small" $\mu$ & Theoretical & $\pm 0.5$\\
$.05\,m_{3/2}\leq \mu \leq .15\,m_{3/2}$ & ${\rm Theoretical + Experimental}$ & $\pm 1.1$\\
\hline
``Large" $\mu$ & Theoretical & $\pm 0.5$\\
$0.5\,m_{3/2}\leq \mu \leq 1.5\,m_{3/2}$ & ${\rm Theoretical + Experimental}$ & $\pm 1.25$\\
\hline
\end{tabular}
\caption{Uncertainties in the calculation of the Higgs mass for a given value of $m_{3/2}$ and $\tan\beta$, as shown in Figure 1. All masses are in GeV.}
\end{table}}

It is important to understand the origin of the spread in the Higgs mass for a given value of $m_{3/2}$ and $\tan\beta$, seen in Figure 1. 
This spread arises from theoretical and experimental uncertainties schematically shown in Table I. 
The two cases in Table I correspond to ``Small" $\mu$ and ``Large" $\mu$ as mentioned in section \ref{Higgs-general}. 
As the name suggests, ``Theoretical"  in the second column corresponds to the variation of input quantities from the theory at the unification scale.
For a given $m_{3/2}$, this includes the variation in the trilinears $A$ and those in the gaugino mass parameters $M_1,M_2,M_3$ consistent with theoretical expectations. 
``Experimental", on the other hand, stands for the variation of the experimental inputs, the top mass $m_t$ and the $SU(3)$ gauge coupling $\alpha_s$, within the current uncertainties.  
The precise variations in the theoretical and experimental inputs are shown in Table II.

{\renewcommand{\arraystretch}{1.5}
\begin{table}[h!]
\begin{tabular}{ | p{3.6 cm} |p{4.8cm}|}
\hline
Theoretical& Experimental\\ 
\hline
 $600 \leq m_{\tilde{g}} \leq$ 1200& $172.2 \leq m_t \leq 174$ \cite{arXiv:1107.5255}\\
  $0.8\,m_{3/2} \leq A_t \leq 1.5\,m_{3/2}$ & $ 0.1177 \leq \alpha^{\overline{MS}}_{s}(m_Z) \leq$ 0.1191 \cite{arXiv:0908.1135}\\
\hline
\end{tabular}
\caption{Variation of the theoretical and experimental inputs. All masses are in GeV.}
\end{table}}
The variations in the bino and wino mass parameters $M_1$ and $M_2$ have a negligible effect on the Higgs mass, and are not shown above. Although we have not fully estimated uncertainties arising from higher-loop effects in the RGE and threshold effects, the fact that our results agree so well with SOFTSUSY suggests that these are at most of the same order as those listed in Table I. Finally, let us discuss the uncertainty in the gravitino mass scale. Figure 1 shows the Higgs mass for three different values of $m_{3/2}$ - 25 TeV, 50 TeV and 100 TeV.  As explained at the end of section \ref{motivation}, the lower limit on $m_{3/2}$ of about 25 TeV arises from the general result that the lightest modulus mass is generically of the same order as $m_{3/2}$. The modulus decays with a decay constant which is effectively suppressed by the string scale, and the requirement of generating a sufficiently high reheat temperature so that BBN occurs in the usual manner, puts a lower bound on $m_{3/2}$. Therefore, the lower limit on $m_{3/2}$ is uncertain only by a small amount. Although the upper limit is less tightly constrained, theoretical expectations constrain it to be not be much above 100 TeV. This is because in the string/$M$ theory vacua considered here, gaugino masses are suppressed only by one-to-two orders of magnitude relative to $m_{3/2}$ in accord with theoretical expectations \cite{hep-ph/0511162,arXiv:0801.0478}. Therefore, the requirement of gauginos to be light enough (with masses $\lesssim$ TeV) such that they 
are part of the low-energy theory at $M_{susy}$ as assumed in section \ref{matching-msusy}, puts an upper limit on $m_{3/2}$ of about 100 TeV. 
A similar upper bound also arises in realistic moduli stabilization mechanisms satisfying the supergravity approximation \cite{arXiv:0801.0478}. 
Improvements in data as well as theory in the future will be extremely helpful in constraining the gravitino mass.

\section{Conclusions}\label{conclude}

Recent progress in string/$M$ theory compactifications which stabilize the moduli and give rise to low-scale supersymmetry  imply that once cosmological constraints are imposed, 
generically the gravitino mass is heavier than about 25 TeV and the scalar masses and trilinears are close to the gravitino mass.  This in turn implies that
the two-doublet Higgs sector of supersymmetric models is a decoupling one, i.e. the physical mass spectrum has 
one light Standard Model-like Higgs boson, with the additional states being heavier, also of order the gravitino mass. 
The resulting effective low scale theory depends only on a few input quantities from which many effects relevant for collider physics can be computed. 
With current understanding, these quantities are constrained by the theory but not yet fully calculable.  
The Higgs mass, in particular, 
depends on these inputs mainly through $\tan \beta$, which is equivalent to a particular combination of the inputs, and to the gravitino mass scale $m_{3/2}$ to a smaller extent. 
The dependence on other combinations of input quantities turns out to be rather mild. 
It is, therefore, possible to calculate the predictions for the observable Higgs boson mass as a function of $\tan\beta$ quite accurately.  
The resulting value holds generically in all corners of four-dimensional string/$M$ theory vacua which satisfy the criteria and assumptions outlined in section \ref{motivation}.

We evaluate $M_h$ by writing the effective four-dimensional theory at the compactification scale, and carrying out the
renormalization group running down to the weak scale, including two-loop effects
in the RGEs and all relevant threshold corrections. This gives an absolute prediction of $M_h$ as a function of $\tan\beta$ to an accuracy of about 2 GeV for a given $m_{3/2}$ when all relevant theoretical and experimental uncertainties are included. The uncertainty mainly arises from the uncertainty in the top mass and the variation of the soft parameters at the unification scale within the theoretically allowed limits.
$\tan \beta$ is not yet accurately calculable from string/$M$ theory, although theoretical arguments suggest that within the framework considered, it should lie in the range from around 2 to 15. Since $\mu$ and $B\mu$ are related by the EWSB constraint, one finds that in these vacua, low $\tan\beta \lesssim 5$ is only possible when $\mu$ is comparable to $m_{3/2}$, while $\tan\beta \gtrsim 5$ is possible for $\mu$ suppressed by one-to-two orders of magnitude relative to $m_{3/2}$. This is an important result as the measurement of the Higgs mass would determine (or constrain) $\tan\beta$ and hence the value of $\mu$.

The dependence on $\tan \beta$ is quite valuable.  If the experimental result
for $M_h$ lies within the range predicted in Figure 1, then the value of $\tan \beta$ is measured 
within this framework, something which is extremely difficult to do by other methods.  Initially the measurement will be of limited
precision but will improve fast as the experimental resolution and the accuracy of
the theory improve. If $M_h$ is greater than about 125 GeV, it puts a
lower limit on $\tan\beta$ which is also quite useful. Depending on the value and the accuracy
with which $M_h$ is measured, it may be possible to draw conclusions about the gravitino
mass and the associated scalar masses and trilinear couplings as well. The knowledge of $\tan\beta$ (and possibly scalar masses and trilinears) 
obtained from the Higgs mass measurement can be used as a consistency check and to look for correlated observables such as 
gluino pair production with enhanced branching ratios to third generation 
final states (top and bottom quarks) \cite{arXiv:0801.0478, arXiv:0901.3367}.

%For smaller tanb the result is unavoidable somewhat sensitive
%to tanb because of the Standard Model D-term $Mz^2 cos(2b)^2$ which decreases for tanb
%less than about 7.  
%Other quantities such as the gravitino mass and the trilinear couplings are
%determined at the factor of two level and do not lead to much variation in Mh.  The
%various dependences of Mh are summarized in Table 1.
%There is one additional current uncertainty in the prediction.  Just as the broken
%electroweak symmetry gives a "D-term" contribution to Mh, if the gauge group of the
%full theory below the string scale is larger than the MSSM group it can contribute
%an additional D-term that can shift the value of Mh.  If the prediction of this paper is correct it tells us not only that it is possible to make %successful predictions from string/$M$ theory, but that the gauge group of the theory is indeed
%that of the MSSM!  If not, some extended gauge groups can be studied to see if they
%give the correct number.

Note that our result is strictly only valid when the matter and gauge spectrum below the compactification scale is 
precisely that of the MSSM and gauginos are suppressed by an order of magnitude or so relative to the scalars. 
As is well known, extended gauge groups can give new $D$-term contributions to the Higgs mass.
Similarly, existence of additional states (such as SM singlets, $SU(2)$ triplets, $SU(3)$ charged states, etc.) with Yukawa couplings 
to the Higgs sector can give rise to new tree-level and radiative contributions to the Higgs mass. Even if the matter 
and gauge content is exactly that of the MSSM but gauginos are not suppressed, 
then the beta function for the Higgs quartic below $M_{susy}$ would 
be different and would lead to a different prediction for the Higgs mass in general. 
Such alternatives can be studied in a straightforward manner if necessary.

If the prediction for the Higgs mass turns out to be correct, it would be an extremely important step forward in relating the string/$M$ theory framework to the real world 
and would open up many opportunities for learning about the string vacuum we live in. 
In addition to learning about $\tan\beta$ and $\mu$ as described earlier, 
it could tell us that the gauge and matter content of Nature is indeed
that of the MSSM! If not, this would imply that one or more of the attractive assumptions in the paper have to be relaxed.

\acknowledgments{We would like to thank Daniel Feldman and Aaron Pierce for helpful discussions and comments on the manuscript. We would also like to thank Bobby Acharya for 
his collaboration on several key ideas that went into this work. P.K. would like to thank MCTP for hospitality. 
The work of G.K., R.L, and B.Z. is supported by the DoE Grant DE-FG-02-95ER40899 and 
and by the MCTP. R.L. is also supported by the String Vacuum Project Grant funded through NSF grant PHY/0917807. The work of PK is supported by the DoE Grant DE-FG02-92ER40699.}

\end{document}